\documentstyle[12pt,epsf]{article}
\def\cO#1{{\cal{O}}\left(#1\right)} 
 
\def\cP{{\cal{P}}} 
 
\def\P{{\mbox{P}}} 
 
\def\frac#1#2{ {{#1} \over {#2} }} 
 
\def\half{\mbox{\small $\frac{1}{2}$}}

\def\VEV#1{\left\langle #1\right\rangle} 
 
\def\ie{\hbox{\rm i.e. }}

\def\Tr{\mbox{\scriptsize Tr}}

\def\fun#1#2{\lower3.6pt\vbox{\baselineskip0pt\lineskip.9pt 
  \ialign{$\mathsurround=0pt#1\hfil##\hfil$\crcr#2\crcr\sim\crcr}}} 

\def\beq{\begin{equation}} 
\def\eeq{\end{equation}} 
\def\beql#1{\begin{equation}\label{#1}} 
\def\beeq{\begin{eqnarray}} 
\def\eeeq{\end{eqnarray}} 
\def\bit{\begin{itemize}} 
\def\eit{\end{itemize}} 
\def\re#1{(\ref{#1})} 
 
\def\de{\delta}

\def\s{\sigma} 
\def\l{\lambda} 
\def\cs{\lambda_2}
\def\cg{\lambda_4}

\def\plb#1#2#3{Phys.\ Lett.\ B {\bf #1} (19#3) #2}

\begin{document}
\setcounter{page}{1}
\begin{flushright}
     IFUM-626-FT \\ 
\end{flushright} 
\par \vskip 10mm 
\begin{center} 
{\Large \bf 
Monte Carlo simulations and field transformation:
the scalar case\footnote{Research supported in part by MURST, Italy}
} 
\end{center} 
\par \vskip 2mm 
\begin{center} 
{\bf 
B.\ All\'es, P.\ Butera, M.\ Della Morte and G.\ Marchesini}\\ 
\vskip 5 mm 
Dipartimento di Fisica, Universit\`a di Milano \\ 
and INFN, Sezione di Milano, Italy 
\end{center} 
 
\par \vskip 2mm 
\begin{center} {\large \bf Abstract} \end{center} 
\begin{quote} 
We describe a new method in lattice field theory to compute observables 
at various values of the parameters $\l_i$ in the action $S[\phi,\l_i]$.
Firstly one performs a single simulation of a ``reference action'' 
$S[\phi^r, \l_i^r]$ with fixed $\l_i^r$.
Then the $\phi^r$-configurations are transformed into those of 
a field $\phi$ distributed according to $S[\phi,\l_i]$, apart from
a ``remainder action'' which enters as a \break weight.
In this way we measure the observables at values of $\l_i$ 
different from $\l_i^r$.
We  study the performance of the algorithm in the case 
of the simplest renormalizable model, namely the $\phi^4$ scalar theory on 
a four dimensional lattice and compare the method with the 
``histogram'' technique of which it is a generalization. 
\end{quote} 

\newpage 
\section{Introduction} 

It is often necessary to study the observables of a given lattice field 
theory over a wide range of values of the parameters 
(coupling constants and masses).  
This is the case, for instance, when mapping out the phase structure
of the model~\cite{phstr} or in problems which require a fine tuning 
of the parameters~\cite{finet}, 
or in the study of the lattice gauge theory beta-function~\cite{SchF}. 
In a numerical approach to these problems one should perform independent 
Monte Carlo simulations for each desired value of the parameters. 
If there are many couplings and masses this method may become 
prohibitively time consuming. 
 
The  so called ``histogram'' technique~\cite{weight} is usually adopted 
to deal with these problems. Suppose that one wants to study observables 
for a lattice theory of a field $\phi$ with action $S[\phi, \l_i]$ for 
various values of the parameters $\l_i$. 
One runs a single Monte Carlo simulation using a reference 
action $S[\phi,\l_i^r]$ with a fixed set of values of the 
parameters $\l_i^r$. 
Then, by reweighting the generated configurations with the factor 
$e^{\Delta S}$, where 
$\Delta S =S[\phi, \l^r_i]-S[\phi, \l_i]$, one computes 
 the observables associated to the action $S[\phi, \l_i]$. 

This method is subject to various limitations~\cite{hmeth,NP}.
 First of all the fluctuations of  $\Delta S$ 
are   of the order of the square root   of the lattice volume.
 Therefore severe statistics problems are met, unless the 
lattice volumes are not too large. More importantly, the statistical 
 significance and the reliability of the results depend on the size of the 
overlap of the regions of the configuration space covered by the importance
 sampling procedure based on the actions $S[\phi,\l_i^r]$ and $S[\phi, \l_i]$
 respectively. As a consequence, the reliability of the results is not 
known a priori~\cite{hmeth,NP}.

We propose a new method to deal with these problems.
 Suppose that one finds a ``field transformation'' from $\phi$, 
distributed according to the action $S[\phi, \l_i]$, 
to a ``reference field'' $\phi^r$, distributed according to 
the ``reference action'' $S[\phi^r,\l^r_i]$. 
More precisely the fields $\phi$ and $\phi^r$ are related by
\beq\label{exact}
D\phi\; \exp \{-S[\phi,\l_i]\}\;\sim\;D\phi^r\;\exp \{-S[\phi^r,\l_i^r]\}
\,,
\eeq
with $D\phi$ the usual (naive) field measure.
Then by a single Monte Carlo simulation with action 
$S[\phi^r,\l_i^r]$ one 
generates $\phi^r$-configurations which can be transformed 
into the importance sampling configurations for $\phi$.

In general, it is difficult to construct exactly the transformation 
\re{exact} and one has to resort to approximations.
In this case one is left with a ``remainder action'' $\de S[\phi^r]$ 
which enters as a factor $e^{\de S}$ in the right hand side of \re{exact}. 
Therefore, also in this approach, one needs to reweight the 
$\phi^r$-configurations generated in the Monte Carlo run.
The efficiency of the method  depends then on: 
i) whether a simple enough field transformation can be  constructed;
ii) how large are the fluctuations introduced by the reweighting from 
the remainder action.
iii) how large is the overlap of the $\phi$-- and of 
the transformed  $\phi^r$--configurations.
The field transformation method is a generalization of the 
``histogram'' technique which, in our language, corresponds
to choose the trivial mapping $\phi^r=\phi$, so that the remainder action
is given by $\Delta S[\phi] = S[\phi,\l_i^r]-S[\phi,\l_i] $.
The advantage of our method is  that one can devise a systematic 
procedure for improving the field transformation by reducing the 
``remainder action'' and therefore by increasing the size of the overlap
 of the $\phi$ and of the transformed $\phi^r$ configurations.

We shall study the efficiency of the method in the case of the 
the simplest renormalizable field theory: the $\phi^4$ model on a 
four dimensional lattice with $L^4$ sites. The action depends on two 
parameters, $\l_2$ and $\l_4$, the coefficients of the quadratic and 
quartic field monomials respectively.
As a first example we shall consider an ultralocal field transformation, 
\ie one which is the same for all lattice sites. 
We shall study the magnetic susceptibility $\chi$ and 
the second moment of the correlation
function $\mu_2$, for various values of the lattice parameters 
 near  the critical line where the 
square mass gap $m^2=8\chi/\mu_2$ vanishes.
We shall consider lattice sizes $L=8$, $L=16$ and $L=20$. 

Particular attention will be devoted to the statistical fluctuations 
coming from the reweighting due to the remainder action.
Since the field transformation method is a generalization of the 
histogram method, the analysis of the errors can be done by following 
the same procedure as in Ref.~\cite{hmeth,NP}.

Comparing the performance of the two methods, 
 we find that, for the region of parameters and lattice sizes considered, 
the field transformation method always gives  accurate values for the 
observables. The histogram method gives 
estimates of  comparable accuracy 
only when the quartic couplings are the same, $\l_4^r=\l_4$.
When this is not the case, the estimated values of the observables are 
not accurate although the remainder action has no large fluctuations.   

In Sect.~2 we describe the method for the scalar model.  
It will be clear that the method can be generalized to other models. 
In Sect.~3 we describe the numerical performance of the simplest  
field transformation.  
In Sect.~4 we comment our results.
 
\section{Description of the field transformation \\ method} 
 
We consider the scalar field theory on a four  
dimensional lattice with  periodic boundary conditions and 
$L^4$ sites. The lattice action is
\begin{eqnarray}
\label{S} 
S[\phi,\l_2,\l_4] & = &
-\sum_{n\mu} \; \half \phi_n\,(\phi_{n+\mu}+\phi_{n-\mu})
\;+\;\sum_n v(\phi_n) \nonumber \\
v(\phi_n)&=&\frac{\l_2+8}{2} \phi^2_n+\frac{\l_4}{4} \phi^4_n \,.
\end{eqnarray}
Only two parameters $\l_2$ and $\l_4$ are needed. 
The normalization of the field is fixed by the kinetic term.  
The field $\phi_n$ is defined on the lattice site at the position 
given by the four-vector $n$ and the sum over $\mu$ extends to four  
dimensions, $\mu=1, \cdots 4$. 

The Green functions, given by the expectation of field polynomials  
$$
\P[\phi]=\phi_{n_1}\;\phi_{n_2} \cdots \phi_{n_k}\,,
$$ 
are defined by
\beq 
\label{VEV} 
\VEV{\P[\phi]}  \equiv  
\frac{1}{Z}\;\int \prod_n d\phi_n\;e^{-S[\phi]}\; \P[\phi]\,, 
\qquad
Z\equiv \int \prod_n d\phi_n\;e^{-S[\phi]} 
\,. 
\eeq 
Our aim is to evaluate the Green functions on a range of values 
of the lattice parameters $\l_i$ by using the information provided 
by a single Monte Carlo simulation with the reference action 
$ S[\phi^r,\l_2^r,\l_4^r] $.
To this end we look for a mapping of the field $\phi_n$ onto  
a reference field $\phi_n^r$ such that \re{exact} is at least approximately
satisfied. The transformation is defined by the matrix 
\beq\label{J} 
J_{nm}=\frac{\partial \phi_n}{\partial \phi_m^r}\,. 
\eeq 
The field $\phi_n$ is a functional of the reference field,  
$\phi_n=\phi_n[\phi^r]$, so that also the monomial $P[\phi]$ 
is a functional of the reference field 
\beq \label{cP}
\cP[\phi^r] \; \equiv \; \phi_{n_1}[\phi^r] \cdots \phi_{n_k}[\phi^r]
\;=\;\P[\;\phi[\phi^r]\;] \,.
\eeq
By changing the fields according to \re{J}, one has
\beq 
\label{measure} 
\prod_n d\phi_n\;e^{-S[\phi,\lambda_i]} 
\;=\;
\prod_n d\phi_n^r\;e^{-S[\phi,\lambda_i]}\;e^{\Tr \ln J} 
\;=\;
\prod_n d\phi_n^r\;e^{-S[\phi^r,\lambda^r_i]}\; 
e^{\de S[\phi^r]}\,, 
\eeq 
where the ``remainder action'' $\de S$, given by 
\beq 
\label{dS} 
\de S[\phi^r] \;\equiv\;
-S[\phi,\lambda_i] \;+\; S[\phi^r,\lambda^r_i] 
 \;+\; \mbox{Tr} \ln J \
\eeq 
allows for the fact that the transformation 
defined in \re{J} may not exactly satisfy \re{exact}.

The Green function  \re{VEV} can be expressed in terms of weighted  
expectations as follows
\beq \label{key}
\VEV{\P[\phi]}\;=\;\frac{\VEV{\;\cP[\phi^r] \>e^{\de S[\phi^r]}}^r} 
                        {\VEV{\;              e^{\de S[\phi^r]}}^r}\,,
\eeq 
where $\VEV{\cdots}^r$ is the expectation value computed with the 
action $S[\phi^r,\lambda^r_i]$.

To compute the Green functions for various values of 
$\l_2$ and $\l_4$, one proceeds as follows:
by performing a single Monte Carlo simulation with $S[\phi^r,\lambda^r_i]$, 
one  generates a (sufficiently large) sequence of $\phi^r$-configurations. 
Then, for each desired value of $\l_2$ and $\l_4$, one computes, 
by using \re{J} and \re{dS}, the values of $\cP[\phi^r]$ and 
$\de S[\phi^r]$ on each configuration.
Averaging over the configurations one obtains the expectation 
values in the numerator and denominator of \re{key} and therefore 
the Green function.
In the following we discuss the form of the matrix $J$ in \re{J} that we have 
considered.

\subsection{Form of the field transformation}

As stated in the introduction, an appropriate field transformation 
should satisfy two conflicting requirements: 
i) it should be easily computable;
ii) it should produce a weight factor $e^{\de S}$ with small fluctuations.

Let us consider, as a first and simplest 
 example, the following ultralocal field 
transformation 
\beq\label{J1}
J_{nm}=\frac{\partial \phi_n}{\partial \phi_m^r}
\;=\; C\> \de_{nm}\> e^{v(\phi_n)-v^r(\phi^r_n)}  \,.
\eeq
where $C$ is a constant and $v$ and $v^r$ are given in \re{S} with 
parameters $\l_i$ and $\l_i^r$ respectively. 
Implementing the initial condition $\phi_n=0$ for  $\phi^r_n=0$ 
the solution of \re{J1} is
\beq\label{J2}
\int_0^{\phi_n} d\phi\;e^{-v(\phi)}\;=\;
C\;\int_0^{\phi^r_n} d\phi\;e^{-v^r(\phi)}\,.
\eeq
Since both fields are non-compact we require that $\phi_n^r\to \infty$ 
as $\phi_n\to \infty$. This fixes the constant $C$ as the ratio of the 
two integrals extended to infinity and defines $\phi_n$ as 
a strictly monotonic function of $\phi_n^r$.

From \re{J1} we  find (neglecting a field independent term)
\beq
\mbox{Tr} \ln J =\sum_n [\>v(\phi_n) -v^r(\phi^r_n)\>] \,,
\eeq
which gives (see \re{dS}) as  remainder action the difference of the 
 kinetic terms
\beq\label{deS}
\de S\;=\; \half \sum_{n\mu} 
\left( \; \phi_n(\;\phi_{n+\mu}+\phi_{n-\mu})\;-\;
\phi^r_n(\;\phi^r_{n+\mu}+\phi^r_{n-\mu}) \; \right)\,.
\eeq
This is the logarithm of the weight to be used in \re{key} for computing 
the Green functions.
It is clear that too large fluctuations of the weight $e^{\de S}$ would 
spoil the efficiency of the method. 

In order to analyze $\de S$ we consider its expansion around small values of
the fields.
From \re{J2} we obtain the following expansion of $\phi_n$
\beq
\phi_n\;=\;C\;\phi^r_n \; \left(1 + \cO{(\phi^{r}_n)^2}\right),
\eeq
 which gives  
\beq\label{dS1}
\de S \;=\; 
\half (C^2-1) \sum_{n\mu} 
\left(\phi^r_n(\;\phi^r_{n+\mu}+\phi^r_{n-\mu}) \right)
+\cO{(\phi^{r }_n)^4}
\eeq
One expects that the first term  quadratic in the fields gives the
largest contribution to the fluctuations. 
Then one may try to improve the field transformation in order to be 
left with a $\de S$ not containing quadratic terms in the fields.  

As already observed, our method reduces to the usual histogram\break
reweighting technique if one chooses the trivial mapping 
$\phi^r_n=\phi_n$. The corresponding remainder action becomes 
the difference of the potential terms 
$\de S=\sum_n[\>v(\phi_n,\l_i)-v(\phi_n,\l_i^r)\>]$. 

\section{First application}
Our aim in this section is to  test how effective is the 
field transformation method by reproducing some standard 
results for the four-dimensional lattice model of the previous section.

By using the simple field transformation defined in \re{J1}
we compute for various values of $\l_i$ the following two 
physical observables: the susceptibility
\beq
\label{chi}
\chi(\l_2,\l_4)\;=\;\frac{1}{L^4}
\sum_{n,m} \VEV{\phi_{n}\; \phi_{m}}\,,
\eeq
and the second moment of the correlation function defined by
\beq
\label{m2}
\mu_2(\l_2,\l_4)\;=\;\frac{1}{L^4}
\sum_{n,m} \> (n-m)^2\> \VEV{\phi_{n}\;\phi_{m}}\,.
\eeq
In terms of these quantities, 
the mass gap squared is defined by $m^2 \equiv 8\chi/\mu_2$. 

There are various points  to be explored in order to test the 
efficiency and accuracy of the field transformation method. 
The remainder action is the crucial quantity to be kept under control. 
Indeed too large fluctuations of $\de S$ will ruin the 
efficiency and the accuracy of the method. 
The issues we consider are the following:
\bit
\item
the main expected challenge is the increase of the number of degrees 
of freedom, namely we might ask whether the quality of the results will 
worsen going to relatively large lattice volumes. 
The remainder action $\de S$ in \re{deS} being a sum over the whole lattice,  
its fluctuations might  grow rapidly with the square root of the volume and  
the method might become inefficient for large lattices;
\item
at a fixed lattice size and for a given reference point $\l_i^r$ 
we should analyze as many values of $\l_i$ as possible, even relatively 
far away from $\l_i^r$. 
Since the (absolute value of the) remainder action $\de S$ increases 
with the differences $|\l_2-\l_2^r|$ and $|\l_4-\l_4^r|$ (see \re{dS1}),
the efficiency and accuracy of the method might  worsen dramatically
by moving too far away from the reference parameters $\l_i^r$;
\item
finally we should test the method for $\l_i$ close to the 
critical line\break
($\l_2^{\mbox{\scriptsize crit}},\l_4^{\mbox{\scriptsize crit}}$)
of vanishing mass gap. 
Here one expects that the remainder action would exhibit
 the large fluctuations 
which are typical of any quantity near a critical point.
\eit

Most computations have been performed on a $8^4$ lattice, which is 
sufficiently small to make a fast exploration of many values 
of $\lambda_i$ possible.
A few computations on $16^4$ and $20^4$ lattices have been performed 
in order to explore the performance of the method as a function
of  the lattice volume.

\begin{figure}
\begin{center}
\hskip -1cm
\leavevmode
\mbox{\epsfig{file=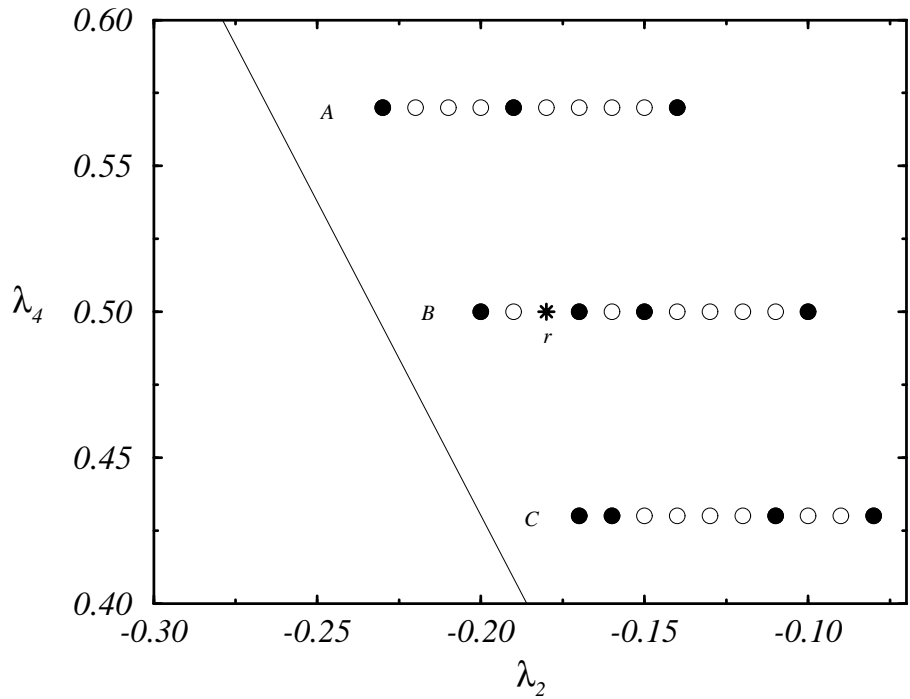,angle=270,width=1.05\textwidth}}
\end{center}
\caption{\it Parameter space. The continuous line 
indicates the one-loop critical line 
 $\cs^{\mbox{\scriptsize crit}} = -0.46479\;\cg^{\mbox{\scriptsize crit}}$.
 The point $r$ denotes the parameters ($\cs^r,\cg^r$)
 of the reference action  used in the Monte Carlo run.
 The performance of the  field transformation method is studied by
 evaluating the observables for  sets of points on the segments 
 $A$, $B$ and $C$ (open and black circles). 
 The black circles represent points at which  Monte Carlo test runs have
 been performed.}
\protect\label{fig:fig1}
\end{figure}

The region of parameters explored in our analysis for the 
$8^4$, $16^4$ and $20^4$ lattices is represented in Fig.~\ref{fig:fig1}.
The reference point $r\equiv (\l_2^r,\l_4^r)=(-0.18,0.5)$, 
represented as a star, indicates the values of 
parameters $\l_2^r,\l_4^r$  used for the reference action. 
The points in the regions denoted by $A$, $B$ and $C$  correspond to 
the values 
of parameters at which we have obtained the observables by using the 
field transformation method. 
At some of these points (black circles)
we also have performed independent Monte Carlo test runs. 
The  continuous line represents the one loop estimate
of the critical line 
($\l_2^{\mbox{\scriptsize crit}},\l_4^{\mbox{\scriptsize crit}}$)
which, in the thermodynamic limit, is associated to the vanishing of 
the mass gap and to the divergence of the susceptibility. 
When approaching the critical points one has ($L\!\to\!\infty$) 
\beq
\mu_2(\l_2,\l_4)\;\sim\;\chi^2(\l_2,\l_4)\to \infty \,, 
\qquad m^2=8\frac{\chi}{\mu_2} \to 0\,,
\eeq
   
\subsection{Field transformation}
A good accuracy is necessary in the calculation of both the constant $C$ 
and the functional dependence of $\phi_n$ on $\phi^r_n$ defined  by \re{J2}. 
This is obtained by  computing the two integrals 
$F(\phi)\equiv\int_0^{\phi} dt\;e^{-v(t)}\;$ and 
$F^r(\phi^r)\equiv\int_0^{\phi^r} dt\;e^{-v^r(t)}$. 
The constant $C$ is obtained by $C=F(\infty)/F^r(\infty)$ and 
is a function of $\l_i$ and $\l_i^r$.  
For small difference of parameters it has the form 
\begin{equation}
C=\left(\frac{\l_4^r}{\l_4}\right)^{1/4} 
  \left[1-\left( \frac{(\l_2+8)^2}{\sqrt{\l_4}} - 
                 \frac{(\l_2^r+8)^2}{\sqrt{\l_4^r}} \right) 
         \frac{\pi \sqrt{2}}{\Gamma\left(1/4\right)^2}\>+\> \cdots 
  \right]
\end{equation}
Then for any value of $\phi^r$, the equation $F(\phi)=CF^r(\phi^r)$ 
is readily solved for $\phi$ and a table of correspondence of values 
 for the mapping ${\phi}^r$ $\rightarrow$ ${\phi}$ 
is created.
This table is then used  every time 
 a measurement is performed for the mapped configuration ${\phi}$.
 The increase in CPU time is negligible compared with the direct 
 Monte Carlo  simulation.

\subsection{Monte Carlo simulation}
We have performed several Monte Carlo simulations with the 
reference action, corresponding to the point $r$ in Fig.~\ref{fig:fig1}, 
on lattices of various sizes.
We have used  the Hybrid Monte Carlo updating algorithm. Each update
consists in 10--15 leapfrog steps of $\Delta\tau = 0.015-0.05$ units of
Langevin time followed by a Metropolis test with $\approx 80\%$ acceptance. 
We discarded approximately 10000 initial sweeps for thermalization. 
The measures were separated by 20--30 decorrelating sweeps.
The errors were evaluated by the usual blocking procedure (see also later). 

We have generated $4\,\times 10^4$  configurations of the 
reference field $\phi^r$ in the thermalized regime. 
For a given $\phi^r$-configuration, by using the field transformation 
\re{J2} from $\l_i^r$ to all considered $\l_i$, we have computed the 
value of the remainder action $\de S_i$ and of 
$$
\tilde \chi_i\;=\;\frac{1}{L^4}\sum_{nm}\;\phi_n\>\phi_m \,,
$$ 
and 
$$
\tilde \mu_{2\;i}\;=\;\frac{1}{L^4}\sum_{nm}\>(n-m)^2\;\phi_n\>\phi_m
\,.
$$ 
 Finally, by taking the appropriate average, we obtain 
 (see \re{key})
\beq
\label{chimu}
\chi(\l_2,\l_4) \simeq \frac{\sum_i \; \tilde \chi_i\;\exp\{ \de S_i\}}
{\sum_i \; \exp\{\de S_i\}}\,,
\qquad
\mu_2(\l_2,\l_4) \simeq \frac{\sum_i \; \tilde \mu_{2\; _i}\;\exp \{\de S_i\}}
{\sum_i \; \exp\{\de S_i\}}\,.
\eeq
for each $\l_2,\l_4$.
Before illustrating the results, let us discuss the fluctuations of $\de S$. 

\subsection{Fluctuation of the remainder action}
\begin{figure}
\begin{center}
\hskip -3cm
\leavevmode
\mbox{\epsfig{file=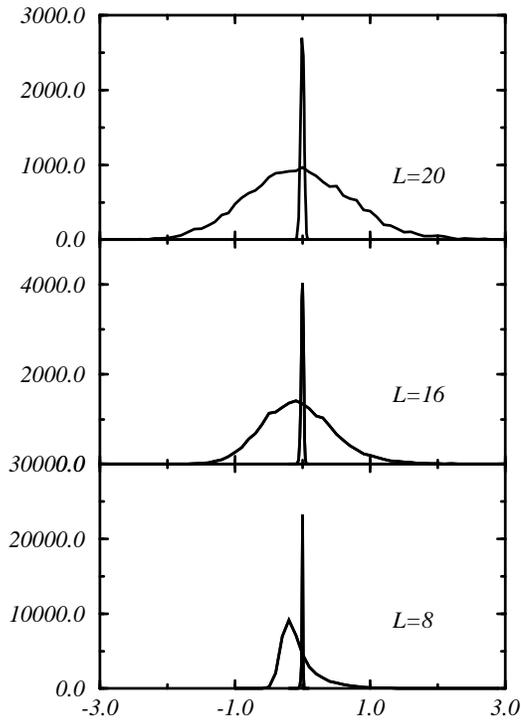,angle=270,width=1.05\textwidth}}
\end{center}
\caption{\it Fluctuations of the remainder action for lattices of size 
 $L\!=\!8,16$ and $20$.
 The field transformation leads from the reference point $r$ 
 (see Fig.~\ref{fig:fig1})
 to the point $\cs\!=\!-0.2$, $\cg\!=\!0.5$.
 The wide histogram corresponds to the remainder action $\de S$ 
 (see \re{deS}).
 The narrow histogram corresponds to $\de' S$, the remainder action minus the
 quadratic term (see  \re{de'S}).
 In this example the quartic coupling is the same, 
 $\cg^r\!=\!\cg$.}
\protect\label{fig:fig2}
\end{figure}

We plot in Fig.~\ref{fig:fig2} (the wider histograms) the distribution of the 
remainder action $\de S_i$ for $8^4$, $16^4$ and $20^4$ lattices 
and for a typical value of $\l_2$ while $\l_4=\l_4^r$.
The fluctuations $\de S_i$ are relatively small and therefore one expects 
that the statistical error of the original Monte Carlo does not 
substantially increase, even for the largest lattice. 
It is expected that the fluctuations grow as the square root of the 
lattice volume. 
Actually we find a slower growth  up to $L=20$. 
The width of the fluctuations for $L=8$, $L=16$ and $L=20$ 
are respectively $0.25$, $0.75$, $1.05$ 
and therefore they appear to increase almost linearly with $L$. 

In view of possible future improvements it is interesting to 
examine which part of the remainder action  gives the largest contribution
 to its fluctuations. 
We  expand $\de S$ in powers of the field 
$\re{dS1}$ and  analyze the effect of the first term. 
In Fig.~\ref{fig:fig2} we have plotted the distribution of the difference 
\beq\label{de'S}
\de' S \;\equiv\; \de S-\half(C^2-1)\>
\sum_{n\mu}\> \phi^r_n\>(\>\phi^r_{n+\mu}\>+\>\phi^r_{n-\mu}\>)\,.
\eeq
We see that the distributions in $\de' S$ are distinctly narrower 
than those of $\de S$ and so we conclude that, in this case, most of 
the ${\de S}$ fluctuations are due to its quadratic part. 
The importance of the quadratic terms in $\de S$ can 
(partially) be understood by observing that the field typically 
assumes small values. For the values of $\l_i$ considered for the 
distributions in Fig.~\ref{fig:fig2}, the values of $\phi^{r\;2}_n$ and 
$\phi^2_n$ are typically of the order of $0.4$. 
Therefore the terms in the expansion \re{dS1} with higher powers in the 
fields are expected give only a small  contribution, both to $\de S$ and  
to its fluctuations.  

\begin{figure}
\begin{center}
\hskip -3cm
\leavevmode
\mbox{\epsfig{file=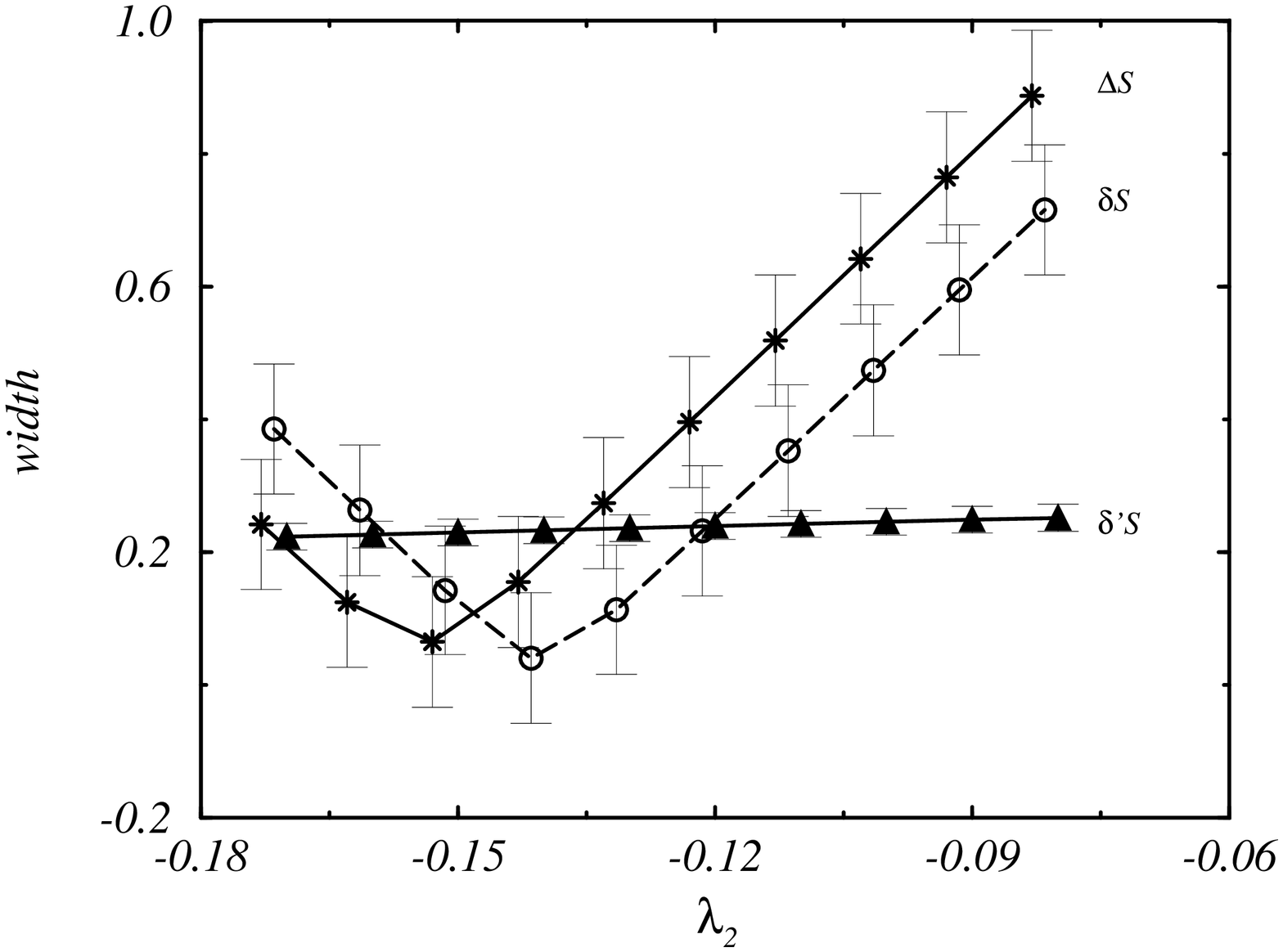,angle=270,width=1.15\textwidth}}
\end{center}
\caption{\it The width of the fluctuations 
of the remainder action as a function 
 of $\cs$ for a $L\!=\!8$ lattice. The reference point is $r$ 
 (see Fig.~\ref{fig:fig1}). 
 The quartic coupling is $\cg\!=\!0.43$ and thus $\cg^r \ne \cg$.
 The open circles correspond to the fluctuations of $\de S$ (see \re{deS}). 
 The triangles stand for the fluctuations of $\de' S$
 (see \re{de'S}). 
 The stars correspond to the fluctuations of 
 $\Delta S=S[\phi,\cs ,\cg]-S[\phi,\cs^r,\cg^r]$ 
 (the logarithm of the weight used in the histogram method). 
 Errors have been estimated by na\"{\i}ve blocking.}
\protect\label{fig:fig3}
\end{figure}

For comparison, we have computed the fluctuations of the remainder 
action of the histogram method which is given by 
$\de S=\sum_n[\>v(\phi_n,\l_i)-v(\phi_n,\l_i^r)\>]$. 
They are similar to those of the field transformation method
as shown in Fig.~\ref{fig:fig3}. Here we have plotted the widths of the
distributions of the remainder action $\de S$ entering in 
the two methods together with the width of $\de' S$ in \re{de'S}.

\subsection{Discussion of errors}
The error analysis in the field transformation method 
is  similar to that for the histogram method~\cite{hmeth,NP}.
Let us assume that, after {\it e.g.} a standard blocking, 
the data are uncorrelated.
First of all, one must take into consideration the correlation between
numerator and denominator in \re{chimu}. 
In general, for an observable ${\cal F}$ given by the ratio 
\begin{equation}
{\cal F} = \frac{\sum_i f_i w_i}{\sum_i w_i} \equiv 
                \frac{\VEV{f w}}{\VEV{g}}\>,
\end{equation}
with $w_i$ a weight, the expression to be used for the propagation 
of errors is 
\begin{equation}
\label{error}
\s_{\cal F}={\cal F}\sqrt{
 \frac{\sigma_{fw}^2}{\VEV{f w}^2} +
 \frac{\sigma_{w}^2}{\VEV{w}^2} - 
  \frac{2\sigma_{fww}^2}{\VEV{f w} \VEV{w}} }\>\>,
\end{equation}
where 
\begin{eqnarray}
 \sigma_{fw}^2 &\equiv& \VEV{f^2 w^2} - \VEV{f w}^2      \nonumber \\
 \sigma_{w}^2 &\equiv& \VEV{w^2} - \VEV{w}^2             \nonumber \\
 \sigma_{fww}^2 &\equiv& \VEV{f w^2} - \VEV{f w} \VEV{w} \nonumber
  \>.
\end{eqnarray}
This expression explicitly shows that large fluctuations in the remainder 
action, giving large values of $\s_w$, would increase the statistical errors
 in the evaluation of the observables.
This fact can be understood also by considering that in this case  
only the few configurations corresponding to the large positive 
fluctuations of $\de S$ would contribute to the observables.

A source of strong systematic error comes from the fact that we have
sampled the Monte Carlo at the reference parameters $\l_i^r$ 
with a finite set of configurations. 
It is known~\cite{NP} that this error goes down with the logarithm
of the number of configurations. 

Finally we recall that in this approach,  measurements 
 at different values of the parameters are 
obtained from the same set of configurations and therefore all of 
them are strongly correlated. 
In practice the correlation matrix of the final measurements is 
almost singular.
This problem is present in any kind of reweighting technique.

\subsection{Results}

\begin{figure}
\begin{center}
\hskip -3cm
\leavevmode
\mbox{\epsfig{file=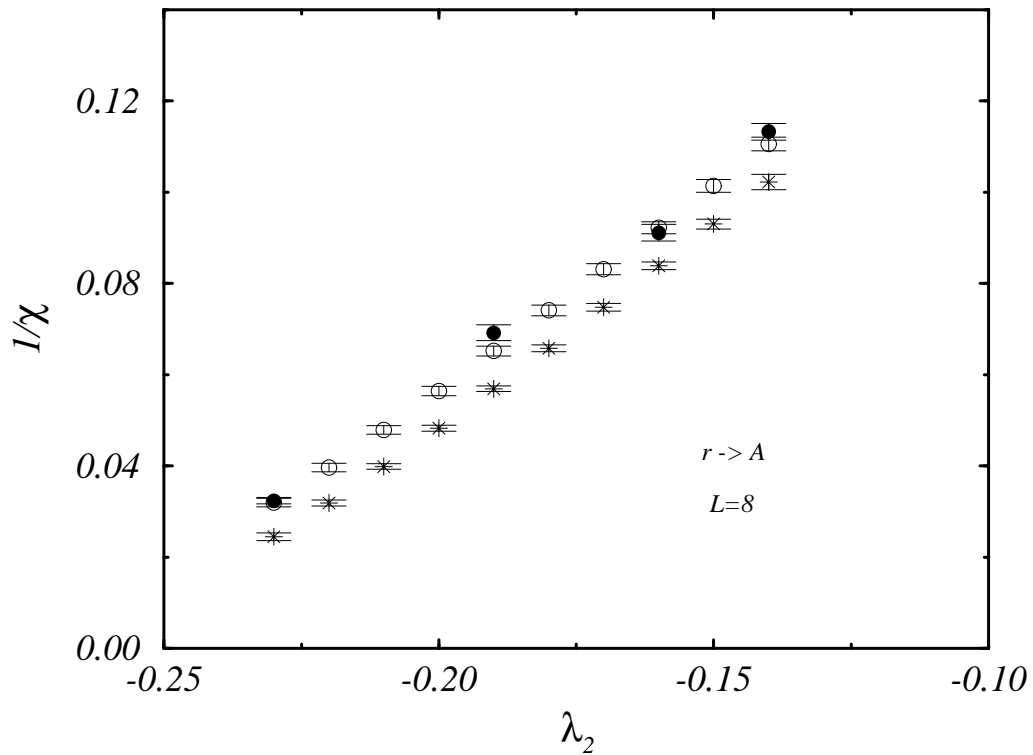,angle=270,width=1.15\textwidth}}
\end{center}
\caption{\it Lattice of size $L\!=\!8$. 
 The emtpy circles represent the values of $1/\chi(\l_2,\l_4)$ 
 obtained by the field transformation  method 
 from the reference point $r$ to  values of 
 ($\cs,\cg$) in the segment $A$ of Fig.~\ref{fig:fig1}.
 The errors are purely statistical and they are computed 
 according to \re{error}.
 We give also the values obtained by Monte Carlo test runs with the
 same statistics (black circles) and the values coming from the
 usual histogram method (stars).}
\protect\label{fig:fig4a}
\end{figure}

\begin{figure}
\begin{center}
\hskip -3cm
\leavevmode
\mbox{\epsfig{file=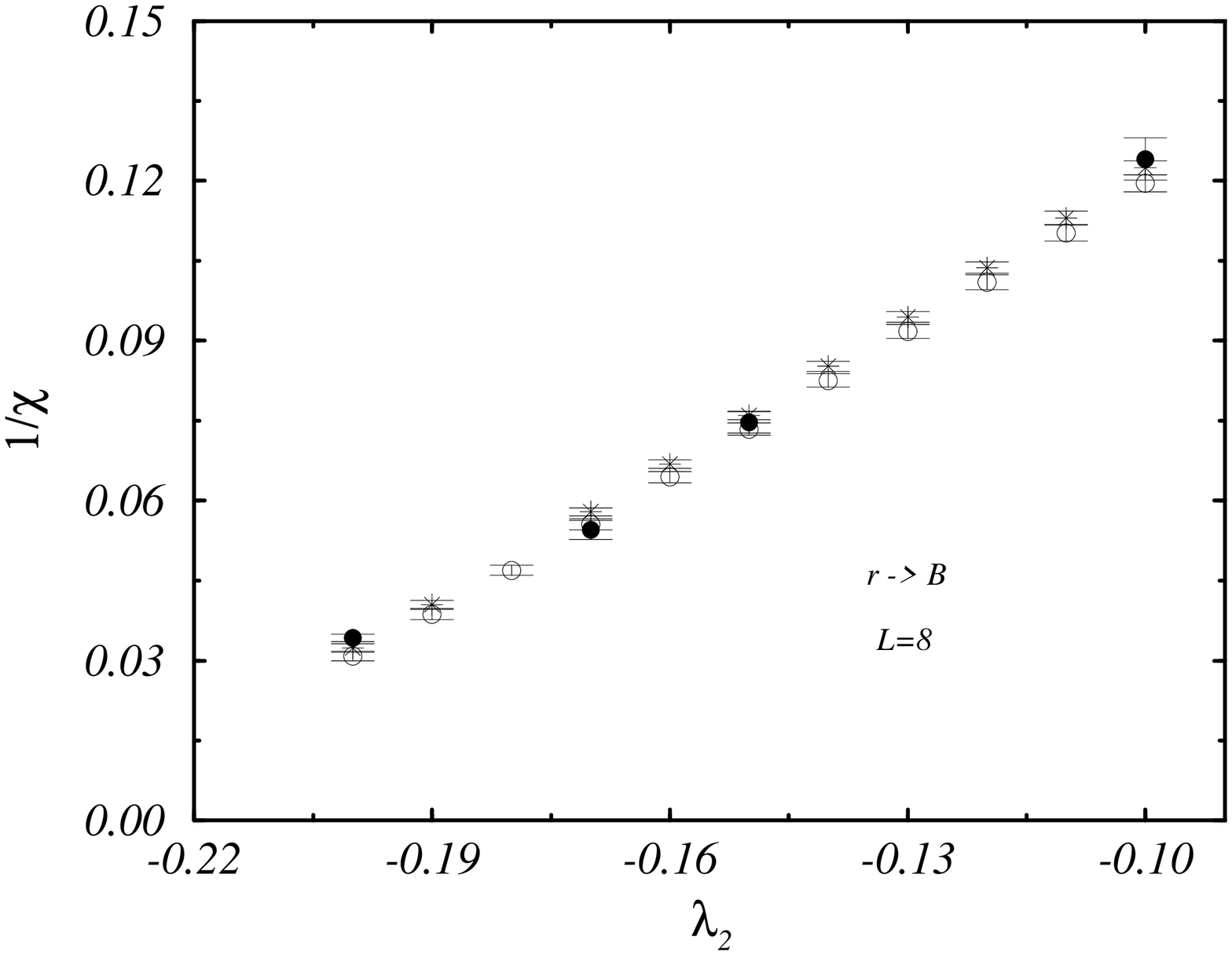,angle=270,width=1.15\textwidth}}
\end{center}
\caption{\it 
 Same as in Fig.~\ref{fig:fig4a}, but for a mapping from 
 the reference point $r$ to  values of
($\cs,\cg$) in the segment $B$ of Fig.~\ref{fig:fig1}.}  
\protect\label{fig:fig4b}
\end{figure}

\begin{figure}
\begin{center}
\hskip -3cm
\leavevmode
\mbox{\epsfig{file=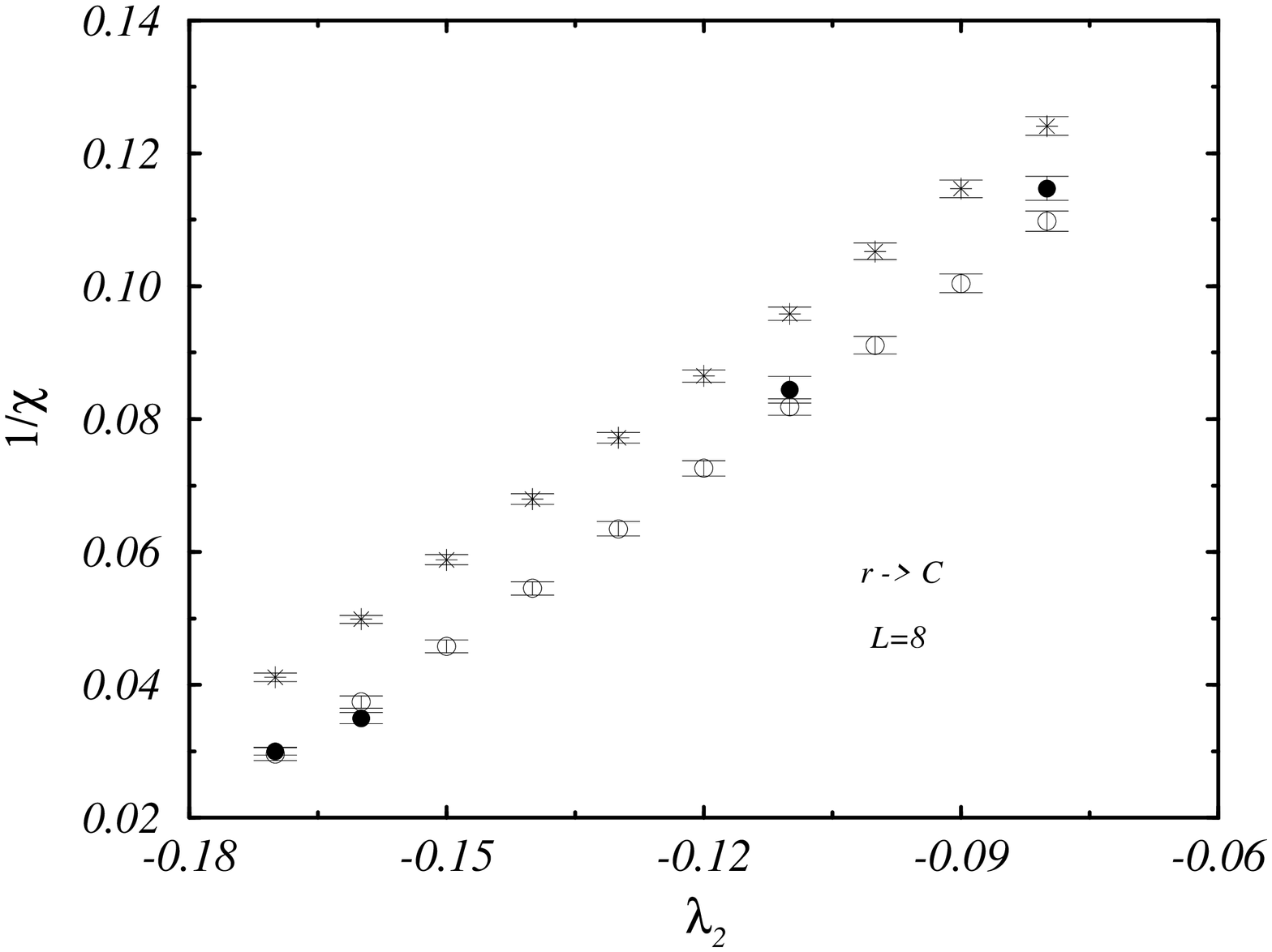,angle=270,width=1.15\textwidth}}
\end{center}
\caption{\it 
 Same as in Fig.~\ref{fig:fig4a}, but for a mapping from 
 the reference point $r$ to  values of
($\cs,\cg$) in the segment $C$ of Fig.~\ref{fig:fig1}.}  
\protect\label{fig:fig4c}
\end{figure}

First we consider the $8^4$ lattice. In 
Figs.~\ref{fig:fig4a}-\ref{fig:fig4c}
we have plotted $1/{\chi(\l_2,\l_4)}$ at the values of parameters 
$\l_i$ corresponding to  points in the three segments 
$A$, $B$ and $C$ (see Fig.~\ref{fig:fig1}). 
The reported errors are obtained using \re{error}.

In order to check the accuracy of the results, we have performed, 
for some values of the parameters, independent Monte Carlo simulations 
with the same number of configurations. 
The corresponding results (black circles) are reported in 
Figs.~\ref{fig:fig4a}-\ref{fig:fig4c} together 
with their statistical errors. 
The agreement between the results obtained by the field transformation 
and by the independent Monte Carlo runs is good even for values of 
the parameters
not very close to the reference point $r$.
We have also computed by the same technique the values of 
$\mu_2(\l_2,\l_4)$ and found similar good agreement. 

In the various plots with fixed $\l_4$ (on the segments $A$, $B$ and $C$) 
the values of $1/{\chi}$ decrease with $\l_2$, but they tend to 
flatten upward before the critical value is reached,
as expected since we work with a finite lattice. 
We also find, as expected, that 
$\mu_2$  is more sensitive to the finite volume effects (see \re{m2}).

The statistical errors obtained by the field transformation method 
are not really larger than those obtained by the direct Monte Carlo. 
This is due to the fact that, as shown in the previous subsection
(see Figs.~\ref{fig:fig2}--\ref{fig:fig3}), 
the $\de S$-fluctuations are not large.

\paragraph{Results for larger volumes.}
Since the remainder action is  a sum over the full lattice
one may expect that the efficiency of the method will worsen by 
increasing the lattice volume. 
However we have found (see Figs.~\ref{fig:fig2}--\ref{fig:fig3}) 
that, even for large lattices, 
the $\de S$-fluctuations are not too large. 
Therefore the agreement with the results from a direct  Monte Carlo test run
should not be sizably degraded.

\begin{figure}
\begin{center}
\hskip -3cm
\leavevmode
\mbox{\epsfig{file=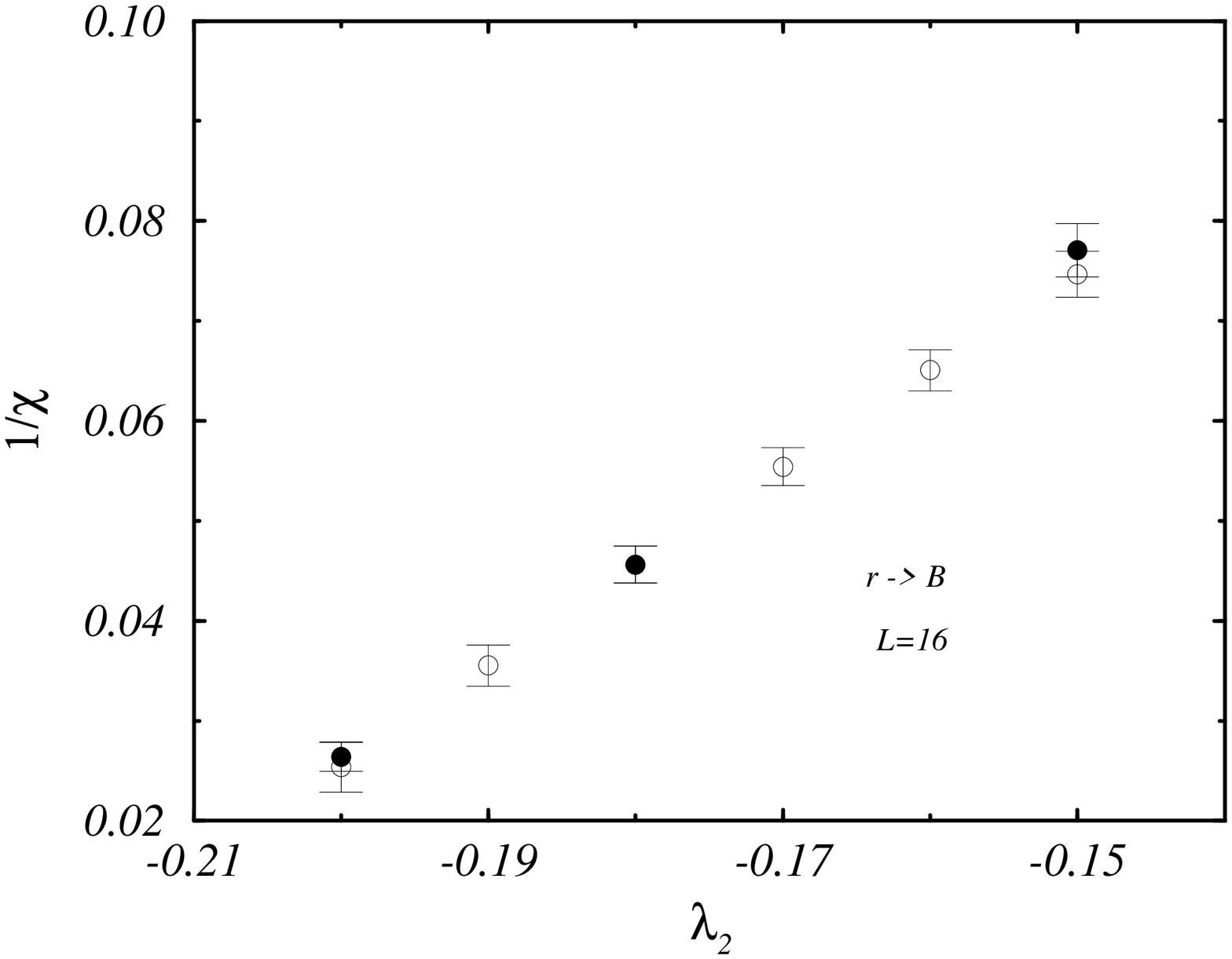,angle=270,width=1.15\textwidth}}
\end{center}
\caption{\it Lattice of size $L\!=\!16$. We have plotted   $1/\chi(\cs,\cg)$ 
 obtained by the 
 field transformation method from the reference point $r$ to parameters
 ($\cs,\cg$)  in the segments $B$ (see in Fig.~\ref{fig:fig1}). 
 The errors are purely statistical and they are
 computed according to \re{error}.
 We give also the values obtained by Monte Carlo test runs with the
 same statistics (black circles). }
\protect\label{fig:fig5}
\end{figure}

\begin{figure}
\begin{center}
\hskip -3cm
\leavevmode
\mbox{\epsfig{file=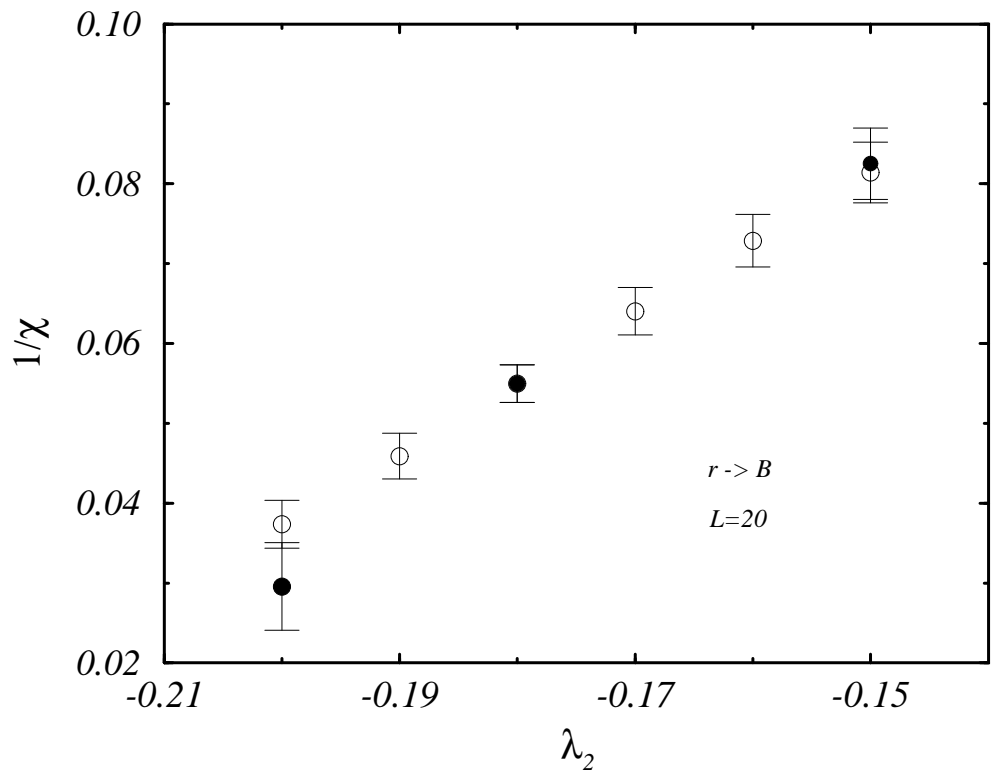,angle=270,width=1.15\textwidth}}
\end{center}
\caption{\it Same as in Fig.~\ref{fig:fig5} for a lattice of size $L\!=\!20$.}
\protect\label{fig:fig6}
\end{figure}

This expectation is confirmed by our study for larger lattices.
In Figs.~\ref{fig:fig5} and \ref{fig:fig6} we have plotted the values of 
${1/ {\chi(\l_2,\l_4)}}$ obtained for $16^4$ and $20^4$  lattices 
respectively. 
 Like in Figs.~\ref{fig:fig4a}-\ref{fig:fig4c}, 
 we have reported  also the measures
 obtained by  Monte Carlo 
test simulations with the same number of independent configurations. 
The agreement with the direct Monte Carlo results is still good. 
 As expected, the statistical errors are larger, since the fluctuations 
of the remainder action increase. However the increase of the 
errors is not dramatic and one still has a good determination of the 
observables.  

\paragraph{Comparison with the histogram method.}
In Figs.~\ref{fig:fig4a}-\ref{fig:fig4c} 
we plot also the results obtained by the 
histogram method (stars). 
We have to distinguish the case in which the quartic coupling $\l_4$
is the same as that of the reference action $\l_4^r$ or is not.
In the first case (see Fig.~\ref{fig:fig4b} in which $\l_4=\l_4^r=0.5$)
we find that the results obtained by the Monte Carlo test simulations
agree with those obtained both by the field transformation and by 
the histogram method.

If instead $\l_4\ne\l_4^r$ (see Figs.~\ref{fig:fig4a} and \ref{fig:fig4c}) 
we find that
the results of the Monte Carlo test simulations
disagree with the ones of the histogram method, while they still
agree with the ones of the field transformation. 
This indicates that for $\l_4\ne\l_4^r$ the regions of field
configuration space probed by the histogram method for the two actions 
$S[\phi,\l_i]$ and $S[\phi,\l_i^r]$ have a small overlap.
Therefore in order to obtain a reliable result by the histogram method
one has to considerably increase the number of configurations sampled. 
The alternative we have proposed is to improve the overlapping 
by a field transformation.

\section{Discussion and conclusions}
The field transformation method, which we have introduced and analysed, 
can be viewed as a generalization of the usual histogram method.
In principle it has the advantage that, by a suitable choice of the 
 mapping, one has the possibility of reducing the remainder 
action, the crucial quantity entering into the weight which can endanger 
the statistical significance of the calculation. 
We have analyzed the case of a simple mapping, the
ultralocal one in \re{J1}.
It gives a remainder action which has  small fluctuations in a wide 
range  of parameters and lattice sizes. 
Moreover the results for the susceptibility (and other computed quantities)
obtained by the field transformation agree quite well with those
obtained by Monte Carlo test runs, even when the lattice volume
increases (at least up to $L\!=\!20$, 
see Figs.~\ref{fig:fig4a}-\ref{fig:fig6}). 
This implies that the region of $\phi^r$-configurations sampled by 
the Boltzmann weight $e^{-S[\phi^r,\l_i^r]}$, when
transformed by \re{J1}, overlaps sufficiently well with the region
probed by $e^{-S[\phi,\l_i]}$.

Since the ultralocal transformation  already gives 
small fluctuations and consistent results, we have not been forced  
to improve it. However we have indicated a possible strategy to
further reduce the remainder action. 
The error analysis and the statistical independence of the results 
obtained by the field transformation method can be discussed 
analogously to the case of the histogram method~\cite{hmeth,NP}.

The histogram method corresponds to the trivial field transformation: 
$\phi_n^r=\phi_n$.
 For $\l_4\ne\l_4^r$, we find that the results of this method disagree 
 with those obtained by  Monte Carlo test runs. This indicates that,  
 for $\l_4\ne\l_4^r$,
the relevant $\phi$-configuration region for the action $S[\phi,\l_i]$ 
is not sufficiently probed by this method.

\vskip 1 cm \noindent
{\bf Acknowledgments}

\noindent
We are grateful to M.\ Bonini, C.\ Destri, E.\ Onofri and M.\ Pepe 
for illuminating discussions.

\end{document}